\documentclass{IEEEtran}	

\usepackage{eqnarray,amssymb,amsmath,amsthm}
\usepackage{mathrsfs}
\usepackage[utf8]{inputenc}
\usepackage[T1]{fontenc}
\usepackage{graphicx}
\usepackage{epsfig}
\usepackage[english]{babel}
\usepackage{subfigure}
\usepackage{color}
\usepackage{epstopdf}
\usepackage{multirow}
\usepackage{cases}
\usepackage{mathtools}
\usepackage{color, colortbl}
\usepackage{cite}
\usepackage{mathtools}
\usepackage{algorithm}
\usepackage{cleveref} % for refereing multiple equations
\crefname{equation}{}{}

\usepackage{algpseudocode}
\theoremstyle{definition}
\newtheorem{defn}{Definition}

\usepackage{algpseudocode}

\begin{document}
\title{Towards Low-Latency and Ultra-Reliable Vehicle-to-Vehicle Communication}

\author{
\IEEEauthorblockN{Muhammad Ikram Ashraf\IEEEauthorrefmark{1}, Chen-Feng Liu\IEEEauthorrefmark{1}, Mehdi Bennis\IEEEauthorrefmark{1}\IEEEauthorrefmark{2}, and Walid Saad\IEEEauthorrefmark{3}}
\\\IEEEauthorblockA{\IEEEauthorrefmark{1}Centre for Wireless Communications, University of Oulu, Finland
\\\IEEEauthorrefmark{2}Department of Computer Engineering, Kyung Hee University, South Korea
\\\IEEEauthorrefmark{3}Wireless@VT, Bradley Department of Electrical and Computer Engineering, Virginia Tech, Blacksburg, VA, USA
\\ E-mail: \{ikram.ashraf, chen-feng.liu, mehdi.bennis\}@oulu.fi, walids@vt.edu}
}

%\IEEEpubid{978--1--5386--3873--6/17/\$31.00~\copyright~2017 IEEE}
\maketitle

\begin{abstract}
Recently vehicle-to-vehicle (V2V) communication emerged as a key enabling technology to ensure traffic safety and other mission-critical applications. In this paper, a novel proximity and quality-of-service (QoS)-aware resource allocation for V2V communication is proposed. The proposed approach exploits the spatial-temporal aspects of vehicles in terms of their physical proximity and traffic demands, to minimize the total transmission power while considering queuing latency and reliability. Due to the overhead caused by frequent information exchange between vehicles and the roadside unit (RSU), the centralized problem is decoupled into two interrelated subproblems. First, a novel RSU-assisted virtual clustering mechanism is proposed to group vehicles in zones based on their physical proximity. Given the vehicles' traffic demands and their QoS requirements, resource blocks are assigned to each zone. Second, leveraging techniques from Lyapunov stochastic optimization, a power minimization solution is proposed for each V2V pair within each zone. Simulation results for a Manhattan model have shown that the proposed scheme outperforms the baseline in terms of average queuing latency reduction up to 97\% and significant improvement in reliability.

% This is done by jointly considering Channel (CSI) and Queue (QSI) State Information when establishing links.

%Recently vehicle-to-vehicle (V2V) communication emerged as a key enabling technology to ensure traffic safety and other mission-critical applications. The main challenges in V2V are latency and reliability, which, to date, remain under-explored.  In this work, we investigate the problem of V2V communication with a clean-slate design in terms of latency and reliability. Specifically,  the network objective is cast as a transmit power minimization subject to latency and reliability constraints which are modeled as  probabilistic constraints of the data queue length. Leveraging techniques from Lyapunov stochastic optimization, we propose a novel hierarchical framework. First, the roadside unit (RSU) dynamically group vehicles into zones and allocates resources based on vehicles' proximity information, traffic demand, and latency desideratum. After acquiring a judicious pool of resources from the RSU, each vehicle pair locally minimizes its transmit power while satisfying the latency-reliability constraints. Simulation results for Manhattan model have shown that the proposed scheme yields significant performance gain in terms of latency reduction up to 97\% and achieve 99.99\% reliability as compared to baseline approach (non-RSU).

%Numerical comparison with a non-RSU scheme shows that in the proposed approach, vehicles achieve 97\% latency reduction and enhance reliability performance thanks to vehicle clustering and resource allocation at the RSU.

\end{abstract}

%\vspace{-0.2cm}
\section{Introduction}
\label{sec:intro}

Vehicle-to-vehicle (V2V) communication is envisioned as one of the most promising enablers for intelligent transportation systems (ITSs) \cite{GA2013}.  Although some applications in V2V communication, e.g.,  lane change alerts and automatic braking, have already been deployed, safety concerns in autonomous transportation and other mission-critical applications still pose significant challenges for vehicular networks.
V2V safety services aim at reducing the risk of traffic accidents. In this regard,  ETSI has standardized two safety messages: cooperative awareness message (CAM) and decentralized environmental notification message (DENM) \cite{cams}. Transmitting these messages with ultra reliability and low latency is a crucial requirement. Although IEEE 802.11p supports V2V communication, it suffers from unbounded latency and varying quality-of-service (QoS) guarantees  \cite{AV2012}. According to 5G requirements, a queuing latency of 0.125\,ms is required  in order to achieve 1\,ms end-to-end latency \cite{Nokia}.
%Queuing latency is one of the major factor in 5G requirements in order to minimize the end-to-end latency ().
%In terms of 5G requirements given by Nokia, the end-to-end latency is 1\,ms with 0.125\,ms queuing latency \cite{Nokia}.

%In terms of requirements, the European project METIS enforces vehicles to transmit small packets with 99.999\% reliability under 5\,ms  latency constraints \cite{METIS}. 

Radio resource management (RRM) plays a key role in the performance of wireless vehicular systems, and it faces many
new challenges in the context of stringent QoS-based V2V communications \cite{MehdiSmallCell}. To address the RRM challenges, a number of recent works have emerged focusing on latency and reliability in V2V communication \cite{ WSun2016, MBotsov2016, SZhang2016}. The authors in \cite{WSun2016} propose a centralized heuristic QoS-based RRM for V2V communication focusing on a sum-rate maximization problem without accounting for the queue state information (QSI). A location-dependent resource allocation by enabling hierarchical clustering is proposed in \cite{MBotsov2016} for V2V communication. Furthermore,  full buffer traffic model is considered, and a subset of orthogonal resource blocks (RBs) are reserved for V2V communication within the same cell \cite{MBotsov2016}.  An RB sharing algorithm which limits the accumulated interference to ensure reliability of V2V communication is proposed in \cite{SZhang2016}. Most of the these works on V2V resource allocation focus on full buffer traffic and do not account for queuing latency, instantaneous traffic demand, and bounded queue size in order to reduce interference. Moreover, this prior art fails to address issues pertaining to the joint optimization of  queuing latency and reliability for V2V-based resource allocation \cite{WSun2016, MBotsov2016, SZhang2016}.

%Prior art addressing reliability and latency issues in V2V communication can be found in  \cite{WSun2016, MBotsov2016, SZhang2016}.
%In these works, reliability constraints are modeled as an outage probability constraint \cite{WSun2016}, signal-to-interference-and-noise ratio (SINR) threshold \cite{MBotsov2016}, and aggregate interference threshold \cite{SZhang2016}.
% {\color{red}Transmission delay is considered in latency measure in those works.} 
%Additionally, due to the randomness of traffic arrival and the time-varying wireless channel, communication capacity might not be able to support the instantaneous traffic demand at all times. In this case, the undelivered data will remain in a queue buffer. When measuring latency, waiting time in the queue should also be accounted for. Nevertheless,  in all aforementioned works, traffic arrivals and data queue length which affect latency performance are ignored.

%  \IEEEpubidadjcol
The main contribution of this paper is a clean-slate design of resource allocation while considering queuing latency and reliability requirements for V2V communication. Unlike previous works such as \cite{WSun2016} and \cite{SZhang2016}, our proposed work jointly considers resource allocation and power control for interference mitigation while blending geographical information and queue dynamics. In order to capture QSI, each vehicle user equipment (VUE)  transmitter uses a queue buffer to store data intended for its VUE receiver.
%Motivated by the mentioned shortcomings, we take into account the proximity-aware information, stochastic data arrival and queue dynamics at vehicle user equipment (VUE).
%Motivated by this shortcoming, we take into account the stochastic data arrival and queue while investigating the V2V latency performance. Different from the SINR-based and other thresholds, we model the reliability constraint as a probabilistic constraint on the queue length.
%In this work, we consider a Manhattan mobility model in which multiple vehicular user equipments (VUEs) under the coverage of a single roadside unit (RSU) engage in V2V communication.
The network problem is formulated as a network transmit power minimization problem subject to probabilistic realibility constraints on vehicle's data queue length. We further utilize Markov's inequality to transform the probabilistic constraint into a latency requirement in terms of average queue length. To solve the formulated problem, we propose a roadside unit (RSU)-assisted approach which decouples the problem into two sub-problems addressed at the RSU and VUE levels, separately. 
In the proposed scheme, the RSU groups pairs of vehicles into virtual zones within which each V2V pair can optimize its  transmit power over the set of allocated RBs while satisfying  latency and reliability requirements.
Due to the localized nature of traffic safety applications, a new virtual zone formation approach is proposed based on the network topology, traffic demand, and latency requirements.
By leveraging Lyapunov optimization techniques, a power minimization is  addressed at the VUE while ensuring queuing latency requirements and reliability constraints. Simulation results validate the effectiveness and performance of the proposed approach as compared to a baseline in terms of queuing latency and high reliability.

%The rest of this paper is organized as follows:  Section \ref{sec:system} presents the system model followed by the problem formulation in Section \ref{sec:problem}. Section \ref{sec:approach} introduced the RSU-aided zone formation empowering resource allocation while satisfying QoS in terms of queuing latency and reliability requirements. Numerical results are presented in Section \ref{sec:num}. Finally, conclusions are drawn in Section \ref{sec:con}.

%\vspace{-0.6cm}
\section{System Model and Problem Formulation}
\label{sec:system}
\subsection{System Model}
%\vspace{-0.2cm}
%
%
%
\begin{figure}[t]
	\centering
	\includegraphics[width=\columnwidth]{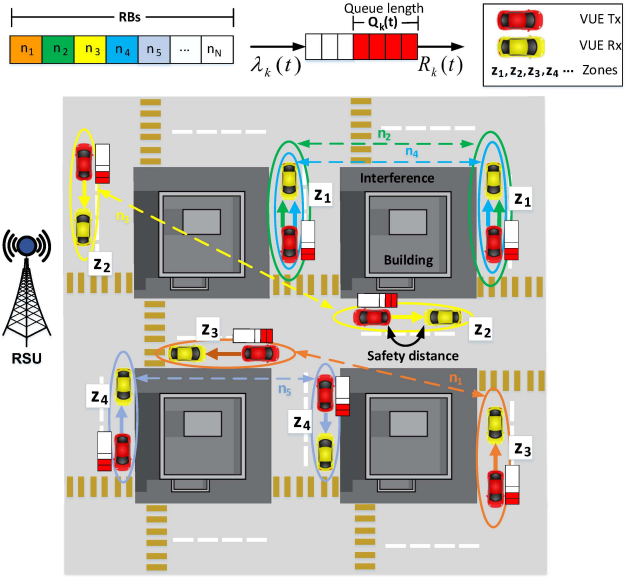}
	%[scale = 0.70]{figures/V2V_zones_v3}
	\vspace{-0.4cm}
	\caption{Considered Manhattan mobility model.}
	\label{fig:zones}
	\vspace{-0.3cm}
\end{figure}
Consider a Manhattan mobility model with V2V communication composed of $K$ VUE transmitter-receiver pairs  under the coverage of a single RSU, as illustrated in Fig.~\ref{fig:zones}. Every VUE pair configuration is fixed during the entire communication period. The safety distance between the transmitter and receiver of a single VUE pair is bounded in a region which is small  compared with the considered network area.
Let $\mathcal{K}$ be the set of all $K$ of VUE pairs who share a set $\mathcal{N}$ of $N$ orthogonal RBs with equal bandwidth $\omega$. Each VUE pair uses multiples RBs whereas one RB can be reused by multiple VUE pairs which causes interference.  Communication  is slotted and indexed by $t$. 
$h_{kk'}^{n}(t)$ denotes the signal propagation channel from the transmitter of pair $k$ to the receiver of pair $k'$ over RB $n$ in slot $t$. Full information of $h_{kk}^{n}(t),\forall\,t,n\in\mathcal{N}$, is assumed to be known at VUE pair $k$.

We define $\eta_k^{n}(t) \in \{0,1\}$ as the RB usage in time slot $t$ where $\eta_k^{n}(t)=1$ indicates that VUE pair $k$ uses RB $n$. Otherwise, $\eta_k^{n}(t)=0$. To send information to the receiver, the transmitter of pair $k$ allocates power $P_{k}^{n}(t)$ over RB $n$ in slot $t$ with $ \sum_{n\in\mathcal{N}}\eta_{k}^{n}(t) P_{k}^{n}(t)\leq P_{k}^{\max}$ where  $P_{k}^{\max}$ is the total power budget.
The data rate of VUE pair $k$ in time slot $t$ is:
\begin{equation*}
%\label{eq:rate}
\textstyle R_{k}(t) = \sum\limits_{n\in\mathcal{N}}  \omega \log_2\Big(1 + \frac{\eta_{k}^{n}(t)P_{k}^{n}(t) {|h_{kk}^{n}(t)|}^2}{\sigma^2 + \sum\limits_{ k' \in \mathcal{K} \setminus k} \eta_{k'}^{n}(t) P_{k'}^{n}(t) {|h_{k'k}^{n} (t)|}^2}\Big),
\end{equation*}
where $\sigma^2$ is the noise variance. Moreover, each VUE transmitter has a queue buffer to store data for its VUE receiver.  Denoting VUE pair $k$'s queue length at the beginning of slot $t$ as $Q_k(t)$, the evolution of $Q_k(t)$ is given by:
\begin{align}
\label{eq:queue-Q}
\hspace{-0.5em}Q_k(t+1) &= \max \big\{Q_k(t) + \lambda_k(t) - \tau R_{k}(t), 0 \big\}.
\end{align}
Here, $\tau$ is the time slot duration while  $\lambda_k(t)$, with  $\bar{\lambda}_k=\mathbb{E}[\lambda_k(t)]$,  is the  traffic arrival at the transmitter of pair $k$ in the beginning of  slot $t$. Our goal is to consider resource allocation and power optimization while satisfying the latency and reliability requirements, in terms of queue length, needed to deliver the safety messages.

\subsection{Problem Formulation}
\label{sec:problem}
Our objective is to find an efficient joint resource and power allocation scheme while meeting the queuing latency and reliability requirements \cite{Nokia}. According to Little's law, the average queuing latency is proportional  to the average queue length \cite{Littlelaw}. Additionally, when the finite-size queue buffer is overloaded, data arrival results in packet drops which incurs unreliable communication. Therefore, the queue length is a key parameter for modeling the queuing latency and reliability for V2V communication. To this end, we impose a probabilistic constraint on the queue length for each VUE pair $k \in \mathcal{K}$, i.e., $\Pr\big(Q_{k}(t)\geq L_{k}\big)\leq \epsilon_{k}$, with an allowable queue length  $L_k$ and a tolerance value $\epsilon_{k}\ll 1$ considered as reliability. In our model, the power consumption for transmission is coupled with the resource usage. Hence, the RSU's objective is to find an optimal network-wide RB usage and power allocation strategy in order to minimize the time-averaged network power while satisfying the probabilistic constraint on queue length. Formally, denoting the network RB usage and  power allocation vectors as $\boldsymbol{\eta}(t)=(\eta_k^n(t),k\in\mathcal{K},n\in\mathcal{N})$ and $\boldsymbol{P}(t)=( P_k^{n}(t),k\in\mathcal{K},n\in\mathcal{N}),\forall\,t$,  respectively, the network-wide optimization problem is written as follows:
\begin{subequations}
	\label{eq:net-obj}
	\begin{IEEEeqnarray}{cl}
	%\begin{align}
	\underset{\boldsymbol{\eta}(t), \boldsymbol{P}(t)}{\mbox{minimize}}&\textstyle~~ \sum\limits_{k \in \mathcal{K}} \sum\limits_{n \in \mathcal{N}}\bar{P}_{k}^{n}  \label{eq:net-obj-1}\\
	\mbox{subject to} 
	&\textstyle~~  \lim\limits_{t\to\infty}\Pr\big(Q_{k}(t)\geq L_{k}\big)\leq \epsilon_{k},~\forall\, k\in\mathcal{K}, \label{eq:net-obj-2} \\
	&\textstyle~~  \sum\limits_{n\in\mathcal{N}}\eta_{k}^{n}(t) P_{k}^{n}(t)\leq P_{k}^{\max},~\forall\,t, k\in\mathcal{K}, \label{eq:net-obj-3} \\
	%&\textstyle~~\sum\limits_{n\in\mathcal{N}} \eta_{k}^{n}(t)\leq N, ~\forall\,t, k\in\mathcal{K},\label{eq:net-obj-4} \\
	&\textstyle~~P_{k}^{n}(t)\geq 0,~\forall\,t,k\in\mathcal{K},n\in\mathcal{N},\label{eq:net-obj-5} \\
	&\textstyle~~ \eta_{k}^{n}(t) \in \{0,1\}, ~\forall\,t, k\in\mathcal{K},n\in \mathcal{N}, \label{eq:net-obj-6} 
	\end{IEEEeqnarray}
	%\end{align}
\end{subequations}
where $\bar{P}_k^{n} = \lim\limits_{T\to\infty}\frac{1}{T}\sum_{t=1}^{T}P_k^{n}(t)$ is the time-averaged power consumption of VUE pair $k$ over RB $n$. 
%Constraint \eqref{eq:net-obj-4} means that each VUE pair $k$ can use all RBs. 
Since it is challenging to analytically analyze \eqref{eq:net-obj-2}, we resort to Markov's inequality to linearize the probabilistic constraint. 
\begin{defn}[\bf{Markov's inequality}]
	\label{def:markov}
	If X is a non-negative random variable and $a > 0$, then
	$ \Pr(X\geq a)\leq \mathbb {E} [X]/a.$
\end{defn}
Using Definition \ref{def:markov}, we  deduce that, given a non-negative queue length, \eqref{eq:net-obj-2} is ensured  if
\begin{equation}\label{eq:Markov_queue}
\textstyle\bar{Q}_{k}= \lim\limits_{T\to\infty}\frac{1}{T}\sum\limits_{t=1}^{T}\mathbb{E}[Q_k(t)]\leq L_{k}\epsilon_{k},~\forall\, k\in\mathcal{K},
\end{equation} 
is satisfied. Note that if  \eqref{eq:Markov_queue} is satisfied, a certain delay performance is guaranteed (by Little's law) since the time-averaged expected queue length will be upper bounded.
Hence, in \eqref{eq:net-obj}, we consider \eqref{eq:Markov_queue} instead of \eqref{eq:net-obj-2}.

\subsection{Lyapunov Optimization Framework}

To find $\boldsymbol{\eta}(t)$ and $\boldsymbol{P}(t)$, we leverage tools from Lyapunov optimization.
%In this section, we will investigate the optimal power strategy and propose a dynamic slot-by-slot power allocation. 
To satisfy \eqref{eq:Markov_queue}, we introduce a virtual queue which evolves as follows:
\begin{equation}\label{eq:queue-F}
F_k(t+1)=\max \big\{F_k(t)+Q_k(t+1) - L_{k}\epsilon_{k}, 0 \big\},~ \forall\,k\in\mathcal{K}.
\end{equation}
Then, the conditional Lyapunov drift-plus-penalty for slot $t$ is given by:
\begin{equation}\label{Eq: Conditional Lyapunov drift}
\textstyle\mathbb{E}\Big[\sum\limits_{k\in\mathcal{K}} \Big(\frac{F_k(t+1)^2}{2}-\frac{F_k(t)^2}{2}+\sum\limits_{n\in\mathcal{N}} VP_{k}^{n} (t)\big)
\big|\mathbf{\Psi}(t)\Big],
\end{equation}
where  $\boldsymbol{\Psi}(t)=(Q_k(t),F_k(t),k\in\mathcal{K})$ for  simplicity.
In \eqref{Eq: Conditional Lyapunov drift}, the non-negative parameter $V$ captures the tradeoff between power cost optimality and latency reduction.
Substituting \eqref{eq:queue-Q} and \eqref{eq:queue-F} into \eqref{Eq: Conditional Lyapunov drift}, and using $(\max\{x,0\})^2\leq x^2$, we obtain 
\begin{multline}
\textstyle\eqref{Eq: Conditional Lyapunov drift}
\leq C+\mathbb{E}\Big[ \sum\limits_{k \in \mathcal{K}}\sum\limits_{n \in \mathcal{N}}VP_k^{n}(t)\\
\textstyle-\sum\limits_{k \in \mathcal{K}}\tau \big( F_k(t)+Q_k(t)+ \lambda_k(t)\big)R_k(t)
\big|\mathbf{\Psi}(t)\Big],\label{Eq: Lyapunov bound}
\end{multline}
where $C$ is a constant. Since this constant does not affect system performance, we omit its explicit expression.
The solution to \eqref{eq:net-obj} can be found by minimizing the upper bound in \eqref{Eq: Lyapunov bound} in each slot, i.e.,
	\begin{IEEEeqnarray}{cl}
%\begin{align}
\underset{\boldsymbol{\eta}(t), \boldsymbol{P}(t)}{\mbox{minimize}} &\textstyle~~ \sum\limits_{k \in \mathcal{K}}\Big[\sum\limits_{n \in \mathcal{N}}VP_k^{n}(t)\notag
\\&\hspace{2em}-\tau\big( F_k(t)+Q_k(t)+ \lambda_k(t)\big)\mathbb{E}[R_k(t)]\big]\label{eq:vue-obj} \\
\mbox{subject to} &~~\eqref{eq:net-obj-3}-\eqref{eq:net-obj-6},\notag
%\cref{eq:net-obj-3, eq:net-obj-4, eq:net-obj-5, eq:net-obj-6}, 
%	&~~\sum\limits_{n\in\mathcal{N}}\eta_{k}^{n}(t) P_{k}^{n}(t)\leq P_{k}^{\max},~\forall\,t, k\in\mathcal{K}, \label{eq:vue-obj-2} \\
%    & ~~\sum\limits_{n\in\mathcal{N}} \eta_{k}^{n}(t)\leq |\mathcal{N}|, ~\forall\,t, k\in\mathcal{K},\label{eq:vue-obj-3} \\
%    &~~P_{k}^{n}(t)\geq 0,~\forall\,k\in\mathcal{K},n\in\mathcal{N},\label{eq:vue-obj-4} \\
%	& ~~\eta_{k}^{n}(t) \in \{0,1\}, ~\forall\, t,k\in\mathcal{K},n\in \mathcal{N}, \label{eq:vue-obj-5} 
%\end{align}
	\end{IEEEeqnarray}
which is an NP-hard mixed-integer programming problem. The expectation is with respect to the interference channel.
Additionally, the RSU needs full global channel state information (CSI) and QSI to solve \eqref{eq:vue-obj} in each time slot. This is clearly impractical for vehicular communication since frequently exchanging fast-varying local information between the RSU and VUE pairs  incurs high overhead. To deal with this, we propose a semi-centralized approach where the RSU is leveraged.

\section{RSU-assisted Resource Allocation}
\label{sec:approach}

The centralized optimization problem in \eqref{eq:vue-obj} is challenging to solve due to the overhead caused by frequent information exchange between VUEs and the RSU at each time slot for optimal RB usage and power allocation. 
In this section, we propose an efficient semi-centralized approach in which we decouple the  problem \eqref{eq:vue-obj} such that the RSU performs RB allocation over a coarse timescale, and the VUE utilizes its local CSI and QSI to optimize the transmit power, at each time slot. For  efficient resource allocation,  the RSU needs to coordinate with the rest of the network at each time slot thus incurring a potentially large information exchange. To this end, $T_0\gg 1$ successive slots are assembled into one time frame to represent the coarse timescale. Hence, vehicles send their local information to the RSU at the beginning of each time frame instead of each time slot $t$. Since severe interference is caused by neighboring vehicles, in order to mitigate the interference among nearby VUEs and meet the QoS requirements as per \eqref{eq:Markov_queue}, the RSU will group vehicle pairs into a set of virtual zones.\footnote{The terms cluster and zone are used interchangeably. Moreover, in vehicle grouping, the VUE transmitter and receiver of one VUE pair are treated as a unity and grouped in the same zone.} In view of this, we let $\mathcal{Z}$ be the set of $Z$ zones. Each zone $z \in  \mathcal{Z}$ is of dynamic size and changes over time frames according to geographical proximity information of VUEs. Here, we let $\mathcal{K}_z$ denote the set of VUEs belonging to zone $z$. Formally, the zone formation rules are defined as:
%Then, we use proportional fair based on the traffic demand and QoS requirements RSU allocates the set of orthogonal RBs 
%, the RSU groups VUE pairs into clusters in which nearby pairs are assigned to different clusters, and RBs are orthogonally allocated to clusters.
%Firstly, to alleviate the overhead caused by frequent information exchange, we assume that the RSU operates in a slow timescale, i.e., every $T_0$ slot with $T_0\gg 1$.  In this regard, $T_0$ successive slots are grouped  into one time frame, and vehicles send their local information to the RSU at the beginning of each time frame.
%Subsequently, we examine the objective function in  \eqref{eq:vue-obj}. If interference in $R_k(t)$ is mitigated, we obtain the solution with the  better objective value. Since severe interference is mainly caused by adjacent vehicles, the RSU groups VUE pairs into clusters in which nearby pairs are assigned to different clusters, and RBs are orthogonally allocated to clusters. In addition, all pairs in a cluster use the same pool of RBs. We further refer to the geographic deployment of a cluster as a \emph{virtual zone}, denoted zone $z$,  in which the set of composed VUE pairs is denoted by $\mathcal{K}_{z}$. 
%The zone formation rules are mathematically defined as
%
%
%
\begin{equation}\label{zoneRule}
\begin{cases}
|\mathcal{K}_z| \geq 1,~\forall\, z \in \mathcal{Z},\\
\mathcal{K}_{z} \cap \mathcal{K}_{z'} = \emptyset,~ \forall \,z,z' \in \mathcal{Z}, z \neq z', \\
\bigcup\limits_{z\in\mathcal{Z}}\mathcal{K}_{z} =\mathcal{K}.\\
\end{cases}
\end{equation}
Each VUE pair  belongs to only one zone whereas each zone has at least one VUE pair. Once the zones are formed, the RSU allocates RBs to zones such that $\eta_z^n\in\{0,1\},\, \eta_z^n=\eta_k^{n},\forall\,k\in\mathcal{K}_z$,  denotes the allocation of RB $n$ to zone $z$. Let $\mathcal{N}_z$ be the set of orthogonal RBs assigned to each zone $z$. In each zone $z$, each VUE $k \in \mathcal{K}_z$ efficiently reuses the allocated RBs and optimizes its power over the given set  $\mathcal{N}_z$ while satisfying  \eqref{eq:Markov_queue}. %Each VUE $k \in \mathcal{K}_z$ efficiently optimize its power over the given set of RBs $\mathcal{N}_z$ while satisfying the latency and reliability constraints.
%Once the zones are formed, RSU initiate RB-allocation mechanism based on the traffic demand and QoS requirement. In our work, propotional fair based  
%Then, we use proportional fair based on the traffic demand and QoS requirements RSU allocates the set of orthogonal RBs 
%Regarding RB allocation at the RSU,  we let $\eta_z^n\in\{0,1\}$ denote the allocation of RB $n$ to zone $z$. Moreover, $\eta_z^n=\eta_k^{n},\forall\,k\in\mathcal{K}_z$.
%For notation clarity, we define the set of allocated RBs  to zone $z$ as $\mathcal{N}_z$.
Note that $\mathcal{K}_z$, $\mathcal{N}_z$, $\eta_z^n$, and $\eta_k^{n},\forall\,z\in\mathcal{Z},n\in\mathcal{N}$, are static during one time frame but dynamic over frames.  Given the  aforementioned zone formation  and RB allocation approaches,  \eqref{eq:vue-obj} can be rewritten as:
\begin{subequations}
	\label{eq:zone_vue_obj}
		\begin{IEEEeqnarray}{cl}
%\begin{align}
	\hspace{-1em}\underset{ \mathcal{K}_z,\boldsymbol{\eta},\boldsymbol{P}(t) }{\mbox{minimize}} &\textstyle~ \sum\limits_{z \in \mathcal{Z}} 
	\sum\limits_{k \in \mathcal{K}_z} \Big[\sum\limits_{n \in \mathcal{N}_z}VP_k^{n}(t)\notag\\ 
	&  \hspace{2em}-\tau\big( F_k(t)+Q_k(t) + \lambda_k(t)\big)\mathbb{E}[ R_k(t)]\big]\label{eq:zone_vue_obj_1}\\
	\hspace{-1em}\mbox{subject to} 
	&\textstyle~~ \sum\limits_{n\in\mathcal{N}_z}P_{k}^{n}(t)\leq P_{k}^{\max},~\forall\,t,k
	\in\mathcal{K}_z,z\in\mathcal{Z}, \label{eq:zone_vue_obj_2} \\
	&  ~~ P_{k}^{n}(t)\geq 0,~\forall\,t,k
	\in\mathcal{K}_z,n\in\mathcal{N}_z,z\in\mathcal{Z},\label{eq:zone_vue_obj_3} \\
	%&~~|\mathcal{K}_z| \geq 1,\; \forall z \in \mathcal{Z},\label{eq:zone_vue_obj_4}\\
	%&~~\mathcal{K}_{z} \cap \mathcal{K}_{z'} = \emptyset, \; \forall z,z' \in \mathcal{Z}, \; z \neq z',\label{eq:zone_vue_obj_5} \\
	%&~~\bigcup\limits_{z\in\mathcal{Z}}\mathcal{K}_{z} =\mathcal{K},\label{eq:zone_vue_obj_6}\\
	& \textstyle~~\sum\limits_{z\in\mathcal{Z}} \eta_z^{n} = 1, ~\forall\, n\in\mathcal{N},  \label{eq:zone_vue_obj_7} \\
	& \textstyle~~\eta_z^n\in\{0,1\},~\forall\, n\in\mathcal{N},z\in\mathcal{Z},  \label{eq:zone_vue_obj_8} \\
	%&\textstyle~~ \sum\limits_{z\in \mathcal{Z}} | \mathcal{N}_z|= N, \label{eq:zone_vue_obj_8}\\
	&~~\mbox{and }\eqref{zoneRule}.\notag
	%\end{align}
		\end{IEEEeqnarray}
\end{subequations}
Next, we present the details  of the proposed zone formation and RB allocation based on QoS requirements (average queue length constraint \eqref{eq:Markov_queue} and mean data arrival $\bar{\lambda}_k$).

\subsection{Proximity-aware Zone Formation and QoS-aware RB Allocation at the RSU}

Due to the spatio-temporal nature of V2V communication, enabling zone formation based on the spatio-temporal proximity of vehicles helps  mitigate the nearby interference within each time frame. Therefore, the RSU performs zone formation while grouping physically distant VUE pairs into a set of zones such that nearby-interference is mitigated.
%In order to reuse the resources within given zone grouping VUE pairs which are physically apart from each other, mitigate the nearby-interference. The key step in zone formation is to identify nearby VUEs and place them in different zone.
%Considering the distance between the transmitter and receiver of each VUE pair is bounded in a small region as  described in Sec.~\ref{sec:system}. Therefore, the middle point coordinate between the transmitter and receiver is representative.  
The geographical locations of the vehicles do not change significantly during one time frame. Therefore, the latest geographical coordinates of each VUE at the beginning of every time frame for zone formation are sufficient. Let $\boldsymbol{v}_k$ and $\boldsymbol{v}_{k'}$ be the geographical coordinates of the VUE pairs $k$ and $k'$, respectively. The Euclidean distance between two pairs $k$ and $k'$ is computed using $d_{kk'} = \Vert\boldsymbol{v}_k - \boldsymbol{v}_{k'}\Vert$. After acquiring all VUE geographical coordinates at the RSU, the following operations are performed to construct zones:
%Additionally,  we assume that  the vehicle's position does not change significantly during one frame time period. 
%The latest coordinate at the beginning of the time frame for zone formation is sufficient. 
%After acquiring all vehicles' geographical coordinates, 
\begin{enumerate}
	\item In order to mitigate near-by interference, RSU randomly selects one VUE pair $k$ and finds other nearest $Z-1$ pairs  with respect to the Euclidean distance.
	\item
	These $Z$ VUE pairs are separately chosen as the first composed elements of $Z$ zones.
	\item
	Subsequently, for unsorted pairs, classify them one-by-one into the farthest zone based on  Euclidean distance.
\end{enumerate}
Algorithm~\ref{Alg: zone} describes the proposed proximity-aware zone formation method.
%
%
%
%Denoting VUE pair $k$'s coordinate and the Euclidean distance between two pairs $k$ and $k'$ as  $\boldsymbol{v}_k$ and $d_{kk'} = \Vert\boldsymbol{v}_k - \boldsymbol{v}_{k'}\Vert$, respectively,
%Denoting VUE pair $k$'s coordinate and the Euclidean distance between two pairs $k$ and $k'$ as  $\boldsymbol{v}_k$ and $d_{kk'} = \Vert\boldsymbol{v}_k - \boldsymbol{v}_{k'}\Vert$, respectively,
%the procedure of distance-based zone formation  is detailed in Algorithm~\ref{Alg: zone}.
%
%
%
Once zones are formed, the RSU focuses on the RB allocation. In order to satisfy the QoS requirement \eqref{eq:Markov_queue}, the VUE pairs with a tighter average queue length constraint $L_k\epsilon_k$ or higher traffic demand $\bar{\lambda}_k$ require more RBs to release the data from their queue buffers. Therefore, considering $L_k\epsilon_k$ and $\bar{\lambda}_k, \forall \, k\in \mathcal{K}_z$, the RSU assigns RBs to zones in a proportional fair manner given by
%\begin{equation}
%\label{eq: prop-fair}
$|\mathcal{N}_z|=\big(\sum_{k\in\mathcal{K}_z}\frac{N\bar{\lambda}_k}{L_k\epsilon_k}\big)/\big(\sum_{z\in\mathcal{Z}}\sum_{k\in\mathcal{K}_z}\frac{\bar{\lambda}_k}{L_k\epsilon_k}\big)$.
%\end{equation}

%Now the zones are formed, we focus on RB allocation.
%To achieve latency requirement \eqref{eq:Markov_queue}, the VUE pair with a tighter average queue length constraint $L_k\epsilon_k$ or higher traffic demand $\bar{\lambda}_k$ needs more RBs to release the data from its queue buffer. Taking $L_k\epsilon_k$ and $\bar{\lambda}_k$  into account,  the RSU allocates RBs to zones by proportional fairness.

\begin{algorithm}[t]
	\caption{Proximity-aware zone formation}
	\begin{algorithmic}[1]
		\Require
		$\boldsymbol{v}_k,\forall\,k\in\mathcal{K}$.
		\Ensure
		$\mathcal{K}_z,\forall\,z\in\mathcal{Z}$.
		\State Initialize $k=1$, $z=1$, $\mathcal{K}_z=\{k\}$, $\tilde{\mathcal{K}}=\mathcal{K}\setminus k$, $\tilde{\mathcal{Z}}=\mathcal{Z}\setminus z$, and $\mathcal{K}_{z'}=\emptyset,\forall\,z'\in\tilde{\mathcal{Z}}$.
		\ForAll{$\tilde{z}\in\tilde{\mathcal{Z}}$}
		\State Find $k^*=\operatorname{argmin}_{k'\in\tilde{\mathcal{K}}} \{d_{kk'} \}$.
		\State Update $\mathcal{K}_{\tilde{z}}=\{k^{*}\}$ and $\tilde{\mathcal{K}}=\tilde{\mathcal{K}}\setminus k^{*}$.
		\EndFor 
		\ForAll{$\tilde{k}\in\tilde{\mathcal{K}}$}
		\State Find $z^*=\operatorname{argmax}_{z\in\mathcal{Z}}\big\{\min_{k'\in \mathcal{K}_z}\{{d_{\tilde{k}k'} }\}\big\}$.
		\State Update $\mathcal{K}_{z^*}=\mathcal{K}_{z^*}\cup \{\tilde{k}\}$.
		\EndFor 
	\end{algorithmic}\label{Alg: zone}
\end{algorithm}

\subsection{Latency and Reliability-aware Power Allocation at the VUE}

\begin{figure}[t]
	\centering
	\vspace{-0.5cm}
	\includegraphics[width=\columnwidth]{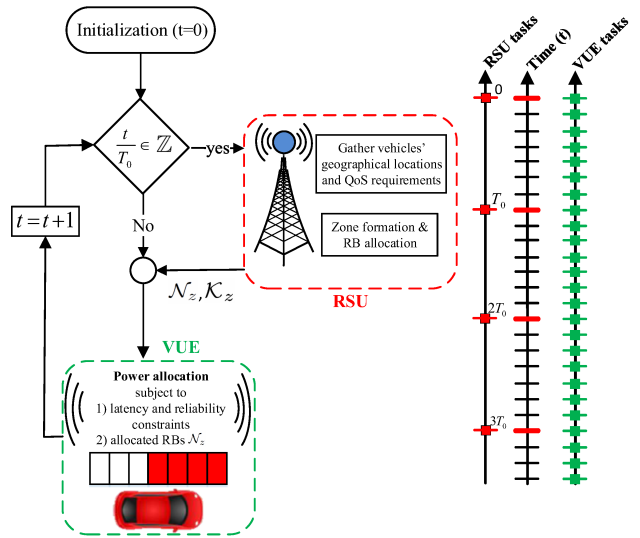}
	%[scale = 0.70]{figures/V2V_zones_v3}
		\vspace{-0.7cm}
	\caption{Flow diagram of the proposed hierarchical resource allocation scheme.}
	\label{fig:flow-diagram}
	\vspace{-0.5cm}
\end{figure}

Once zone formation and RB allocation (i.e., \eqref{zoneRule} and \eqref{eq:zone_vue_obj_7}) are done, each VUE $k\in\mathcal{K}_z$ solves \eqref{eq:zone_vue_obj} locally
%we  decouple \eqref{eq:zone_vue_obj} into $K$ sub-problems which can be solved locally. For the VUE pair $k\in\mathcal{K}_z$ in zone $z$, the decoupled problem is written as follows:
%
%
%
\begin{subequations}
	\label{VUE power}
	\begin{IEEEeqnarray}{cl}
	%	\begin{align}
	\underset{P_{k}^{n}(t)}{\mbox{minimize}}
	&\textstyle~~\sum\limits_{n\in\mathcal{N}_z} VP_{k}^{n}(t)-\sum\limits_{n\in\mathcal{N}_z} \big(F_k(t)+Q_k(t)+ \lambda_k(t)\big) \notag
	\\&\hspace{6em}\textstyle \times \omega \tau \log_2\Big(1 +\frac{P_{k}^{n}(t)h_{kk}^{n}(t)}{\sigma^2 }\Big)\label{VUE power-1}\\
	\mbox{subject to} 
	&\textstyle~~\sum\limits_{n\in\mathcal{N}_z} P_{k}^{n}(t)\leq P_{k}^{\max}, \label{VUE power-2} \\
	&~~P_{k}^{n}(t)\geq 0,~\forall\,n\in\mathcal{N}_z, \label{VUE power-3}
	%\end{align}
\end{IEEEeqnarray}
\end{subequations}
which must be solved at each time slot $t$. Here, the aggregate interference is negligible since orthogonal RBs are allocated to adjacent VUE pairs. Applying the Karush-Kuhn-Tucker (KKT) conditions to the convex optimization problem \eqref{VUE power}, the VUE transmitter finds optimal the transmit power $P_{k}^{n*}(t),\forall\,n\in\mathcal{N}_z$, which  satisfies
\begin{equation}\label{eq:optimal power}
\frac{\big(F_k(t)+Q_k(t) + \lambda_k(t)\big) \omega \tau h_{kk}^{n}(t) }{\big(\sigma^2  +P_{k}^{n*}(t)h_{kk}^{n}(t)\big)\ln 2}
=V+\gamma
\end{equation}
if $\frac{(F_k(t)+Q_k(t)+ \lambda_k(t)) \omega \tau h_{kk}^{n} (t)}{\sigma^2 \ln 2}>V+\gamma$. Otherwise, $P_{k}^{n*}(t)=0$. Moreover, the Lagrange multiplier $\gamma$ is 0 if $\sum_{n\in\mathcal{N}_z} P_{k}^{n*}(t)< P_{k}^{\max}$, and we have $\sum_{n\in\mathcal{N}_z} P_{k}^{n*}(t)= P_{k}^{\max}$ when $\gamma>0$.
%
%
%
%When $F_k(t)$,  $Q_k(t)$, or $\lambda_k(t)$ is large, the  reliability \eqref{eq:net-obj-2} and queuing latency requirements \eqref{eq:Markov_queue} can be easily violated due to large queue length and data arrival (as per \eqref{eq:queue-Q} and \eqref {eq:queue-F}). Therefore, in order to satisfy \eqref{eq:net-obj-2} and \eqref{eq:Markov_queue}, each VUE optimizes the transmit power while focusing on the data rate. Otherwise, power consumption is minimized. 
After sending information, VUE pair $k$  updates $Q_{k}(t+1)$  and $F_k(t+1)$  as per \eqref{eq:queue-Q} and \eqref{eq:queue-F}. The detailed steps of the proposed framework are summarized in Fig.~\ref{fig:flow-diagram}.
%In this case,  the vehicle  increases the data rate by transmitting with more power. Otherwise, the vehicle puts more attention on power cost minimization.

%\vspace{-0.4cm}	
\section{Numerical Results}
\label{sec:num}

In this section, we simulate a Manhattan mobility model with different vehicle densities  that are uniformly distributed over a $250\times 250\,\mbox{m}^2$ area. As shown in Fig.~\ref{fig:zones}, four buildings with 100\,m breadth are deployed and spaced by two lanes for bi-directional traffic mobility. Vehicles travel  along defined roads with an average speed of 50\,km/h. The safety distance between the VUE transmitter and VUE receiver dynamically ranges from 10\,m to 20\,m. For V2V communication, we consider the Berg recursive  path loss  model \cite{METIS} and  line-of-sight propagation within each VUE pair. For all VUE pairs  $k\in\mathcal{K}$, we set $P^{\max}_k=10$\,dBm, $\bar{\lambda}_k=200$\,kbps, $L_k=2000$\,bits,  and $\epsilon_k=0.1$. The other parameters are: $N=15$ RBs, $\omega=180$\,kHz, $\tau=1$\,ms, $\sigma^2=-80$\,dBm, $T_0=100$, and $Z=5$ zones. For the baseline, we refer to the configuration 1 in 3GPP \cite{V2V3gpp}. Therein, each VUE  optimizes its transmit power over all RBs in each time slot.
%Referring to the idea proposed in 3GPP configuration 1\cite{V2V3gpp}, we consider the scheme in which each VUE pair optimizes its power over all RBs as the baseline. %In order to compare the proposed approach with the idea proposed in 3GPP as baseline \cite{V2V3gpp} (configuration 1, distributed resource allocation), each VUE pair optimize its power over the all RBs in every time slot.
%(i.e., configuration 1 for distributed resource allocation) such that each VUE pair can use the whole set of RBs in every time slot.
We use $Q(t)/\lambda$ as the instantaneous queuing latency metric \cite{EstDelay}.
%, and average queuing latency is expressed as  $\bar{Q}/\bar{\lambda}$ using \cite{Littlelaw}.  

\begin{figure}[t]
	\centering
	\vspace{-0.5cm}
	\includegraphics[width=\columnwidth]{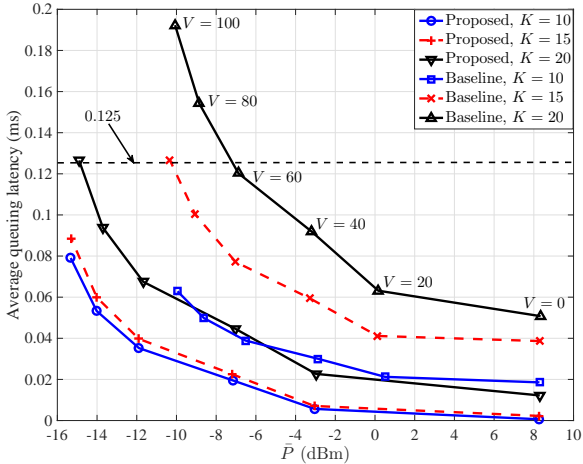}
	\vspace{-0.8cm}
	\caption{Tradeoff between average queuing latency and average power for various densities of VUEs.}
	\vspace{-0.4cm}
	\label{fig:delayVspower}
\end{figure}

%Thus, we have the reliability constrain $\Pr(\mbox{Latency}>10\,\mbox{ms})\leq 0.1$ and 1\,ms average latency requirement  as per  \eqref{eq:net-obj-2}  and \eqref{eq:Markov_queue}, respectively.
The tradeoff between the average queuing latency and the average power consumption is studied in Fig.~\ref{fig:delayVspower} for different densities of VUE pairs.  When $V$ is small as per \eqref{VUE power-1}, the VUE focuses on the rate maximization to decrease its queuing latency which consumes more power. Oppositely, for a large $V,$ the VUE saves  power consumption by allowing the queue length to grow. Fig.~\ref{fig:delayVspower} shows that the proposed approach achieves not only a queuing latency requirements of $0.125$\,ms \cite{Nokia} but also an average queuing latency reduction of up to 97\%  as compared to the baseline for different vehicle densities. Furthermore, the average latency enhancement is more prominent when $K>N$ due to the RSU-aided interference mitigation.

%. From Fig.~\ref{fig:delayVspower}, we can see that the proposed approach achieves up to 97\% average latency reduction. 
\begin{figure}[t]
	\centering
	\includegraphics[width=\columnwidth]{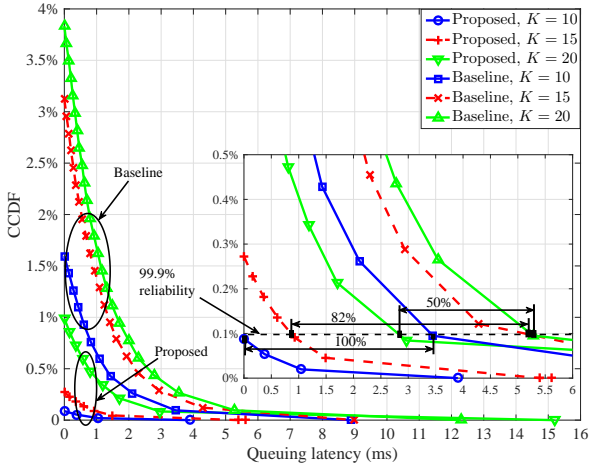}
	\vspace{-0.8cm}
	\caption{CCDF of the instantaneous queuing latency for various densities of VUEs with $V=0$.}
	\vspace{-0.4cm}
	\label{fig:queue99Vspower}
\end{figure}
\begin{figure}[t]
	\centering
	\vspace{-0.5cm}
	\includegraphics[width=\columnwidth]{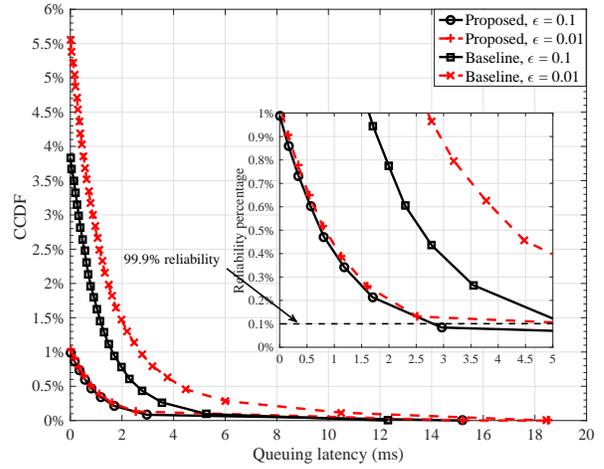}
	\vspace{-0.8cm}
	\caption{CCDF of the instantaneous queuing latency for different $\epsilon$ with $K=20$ VUE pairs and $V=0$.}
	\vspace{-0.4cm}
	\label{fig:cdfEpsilon}
\end{figure}

Subsequently, considering  the case $V=0$ which has the lowest queuing latency, we investigate the transmission reliability via the complementary cumulative distribution function (CCDF) of the instantaneous queuing latency.
The reliability performance for  varying VUE densities and different reliability constraint values $\epsilon$ is shown in Fig.~\ref{fig:queue99Vspower} and Fig.~\ref{fig:cdfEpsilon}, respectively.
It can be seen that, in different network settings, our approach always satisfies the aforementioned reliability constraint as per \eqref{eq:net-obj-2} and achieves higher reliability performance (i.e., lower CCDF values) compared with the baseline. Further, for the 99.9\% reliability case, i.e., 0.1\% of the CCDF value, in Fig.~\ref{fig:queue99Vspower} and Fig.~\ref{fig:cdfEpsilon}, our proposed approach achieves a queuing latency reduction up to 100\%, 82\%, and 50\% for $K$ = 10, 15, and 20, respectively, as compared to the baseline. Finally,  Fig.~\ref{fig:cdfEpsilon} shows that, in order to achieve 99.9\% reliability for different values of $\epsilon$, the proposed approach reaches up to 73\% and 50\% for $\epsilon$ = 0.01 and 0.1, respectively, as compared to the baseline.

\section{Conclusions}
\label{sec:con}
%Taking into account the randomness of traffic arrivals and queue dynamics from a queuing latency and reliability standpoint, 
In this paper, we have studied the performance of V2V communication in terms of queuing latency and reliability. The network objective is to minimize VUEs' power consumption subject to queuing latency and reliability constraints  modeled as probability constraints on data queue length. We have proposed a  novel two-timescale resource allocation approach. In the slow timescale, the RSU forms virtual zones and allocates RBs to zones based on vehicles' spatial information and QoS requirements. Using Lyapunov optimization, every vehicle locally optimizes its transmit power while satisfying queuing latency and reliability constraints. Simulation results have shown that our proposed scheme outperforms the baseline with an average queuing latency reduction up to 97\% and achieves significant improvement in terms of reliability.	

%Simulation results for a Manhattan model have shown that the proposed scheme outperform the baseline approach in terms of average queuing latency reduction and reliability.
	
\section*{Acknowledgments}
This research was supported by the Academy of Finland project CARMA, the NOKIA donation project FOGGY,  the Thule Institute strategic project SAFARI, and the U.S. Office of Naval Research (ONR) under Grant N00014-15-1-2709.

\bibliographystyle{IEEEtran}
\bibliography{ref_eucnc}		 
\end{document}